\documentclass[11pt]{report}

\usepackage[final]{pdfpages}
\usepackage{ucs}
\usepackage{subfigure}
\usepackage{color,graphicx}
\usepackage{rotating}
\usepackage{amsmath}
\usepackage{amssymb}
\usepackage{amsxtra}
\usepackage{wrapfig}
\usepackage{setspace}
\usepackage{version}
\usepackage{fullpage}
\usepackage{framed}
\usepackage[color]{changebar}
\cbcolor{blue}
\usepackage{program}
\usepackage[authoryear]{natbib}

\begin{document}
\title{oligomets filter flow paper}
\begin{flushleft}

\Large

{\bf A filter-flow perspective of hematogenous metastasis offers a non-genetic paradigm for personalized cancer therapy}

\normalsize
Jacob G.\ Scott$^{1,2}$, Alexander G. Fletcher$^2$, Philip K. Maini$^2$, Alexander R.A.\ Anderson$^1$, \& Philip Gerlee$^{1,3}$ \\

$^1$ Integrated Mathematical Oncology, H. Lee Moffitt Cancer Center and Research Institute, Tampa, FL, USA\\
$^2$ Wolfson Centre for Mathematical Biology, Mathematical Institute, Radcliffe Observatory Quarter, Oxford University, Woodstock Road, Oxford OX2 6GG, UK\\
%$^2$ Center for Mathematical Biology, Oxford University, UK\\
%$^3$ Sahlgrenska Cancer Center, University of Gothenburg; Box 425, SE-41530 Gothenburg, Sweden \\
$^3$ Mathematical Sciences, University of Gothenburg and Chalmers University of Technology, SE-41296 Gothenburg, Sweden \\

\bigskip

\textbf{Prepublication draft.}

\vfill
Running title: filter-flow perspective
Correspondence:\\
Jacob G. Scott, jacob.g.scott@gmail.com and Philip Gerlee, gerlee@chalmers.se
\end{flushleft}
\newpage

%\section*{Letter}
\section*{Abstract}
Research into mechanisms of hematogenous metastasis has largely become genetic in focus, attempting to understand the molecular basis of `seed-soil' relationships. Preceeding this biological mechanism is the physical process of dissemination of circulating tumour cells (CTCs). We utilize a `filter-flow' paradigm to show that assumptions about CTC dynamics strongly affect metastatic efficiency: without data on CTC dynamics, any attempt to predict metastatic spread in individual patients is impossible.

%highlight non-genetic opportunities to intervene in the metastatic process, and show that 

\section*{Brief Communication}
\noindent Nearly 150 years after Ashworth's discovery of the vector of hematogenous metastatic disease, the circulating tumor cell (CTC) \cite{Ashworth}, the mechanisms driving this process remain poorly understood and unstoppable \cite{Plaks}.%Watson01012013 
%Since then, many attempts have been made to account for the patterns of metastatic spread. 
For over a century the dominant paradigm has been the seminal, yet qualitative, seed-soil hypothesis proposed by Paget in 1889 \cite{Paget:1989ys}.  This began to be challenged in 1992, when a quantification of the contribution of mechanical and seed-soil effects was attempted by Weiss \citep{weissmemoriam},
%.  As a first attempt to understand the importance of seed-soil effects, Weiss 
who considered the `metastatic efficiency index' (MEI) of individual primary tumors and metastatic sites \cite{Weiss1992}.
%Studying metastasis in a subset of organs, 
He calculated MEI as the ratio of metastatic involvement to blood flow through an organ and
%found that on a logarithmic scale, the MEIs could be separated into 
three classes emerged: low, where the soil-organ relationship is hostile; high, where it is friendly; and medium, where blood flow patterns to a large extent explain patterns of spread. The utility of Weiss' classification method largely ended there, and has since been put aside in favor of genetic correlations \cite{Minn:2005ia,Bos:2009jl}. While illuminating, this approach has yet to offer any actionable conclusions, and its applicability is threatened by the growing understanding that genetic heterogeneity, not clonality, is the rule in cancer \cite{Marusyk:2012uq,Gerlinger:2012uq}. %,Navin:2011kx, 

\begin{figure}[!htb]
\begin{center}
\includegraphics[width=11cm]{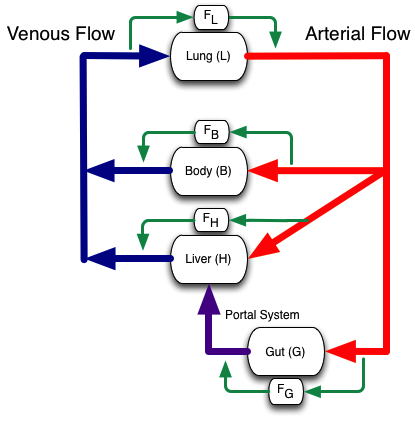}% Here is how to import EPS art
\caption{\label{fig:filterflow}Schematic of the human vascular system network topology.  It is evident by inspection of the network diagram that tumors originating in the gut and lung experience significantly different flow patterns and a different order in which they experience filtration at capillary beds than tumors originating in other parts of the `body' \citep{Scott:2012p1778}. The alternate pathways (green) represent the fraction of cells which evade arrest (filtration) at a given capillary bed.  There are scant measurements of this fraction in the literature, and none in clinical studies that evaluate outcomes. We postulate that by ascertaining the distribution of CTCs in this network for individual patients, information about the existence of subclinical metastatic disease, and therefore metastatic propensity, will come to light, and allow for better staging, prognostication and rational use of organ-directed therapy in the setting of oligometastatic disease.} 
\end{center}
\end{figure}

We seek here to use Weiss' MEI and a recently proposed model of CTC dynamics (summarised in Figure \ref{fig:filterflow}) \cite{Scott:2012p1778,Scott:2013p1914} to extend our understanding of metastasis.
%from a physical perspective.  
In so doing, we can explain, using a \textit{translatable, patient parameter-specific method}, the population-level data for \textit{any} given cancer. Our method also presents a way to utilize `personalized' patient CTC measurements to assay for the burden and distribution of metastatic disease which can be used for guiding organ-directed therapy and more precise staging.

Until recently, even perfect information about the existence and distribution of metastatic disease would have done little to affect treatment choice, as the options were largely limited to the use of systemic chemotherapy.  However, recent years have witnessed the advent of more effective localized therapies for metastatic involvement, in the form of liver-directed therapy, bone-seeking radionuclides and stereotactic body radiation therapy.  These novel modalities have allowed for targeted therapy to specific parts of the body with minimal side efffects and high eradication potential.  Further, trials offering treatment with curative intent to patients with limited, `oligometastatic' disease have shown promise \cite{Milano:2009zr,Milano:2012ys}, although it is not yet possible to identify such patients in an objective manner \cite{Weichselbaum:2011uq}.% \cite{Hellman:1995fk}.  

In this letter we extend Weiss' method to understand metastatic proclivities of certain organ sites and to provide a framework by which to understand the metastatic distribution from novel CTC measurements.  To do this, we consider blood flow between organs \cite{Williams:1989uq}, filtration in capillary beds (see Figure \ref{fig:filterflow} and Table \ref{table:tab1}) and distribution of metastatic involvement in a series of untreated patients at autopsy \cite{Disibio2008}.  For each organ-organ pair we calculate the MEI by normalising incidence by putative CTC flow between the two organs, taking into account the reduction that occurs in capillary beds \cite{Scott:2012p1778,Scott:2013p1914}; which has been shown to be of the order of $10^{-4}$ cell$^{-1}$ \cite{Okumura:2009vn}. This post-capillary bed reduction in CTC numbers can be altered by the presence of micrometastases, which can amplify CTC numbers downstream of their location through shedding. Thus, by adjusting filtration rates throughout the network, we can represent any different configuration of metastatic disease and thus capture different organ-organ metastatic efficiencies.

\begin{figure}[!htb]
\begin{center}
\includegraphics[width=17cm]{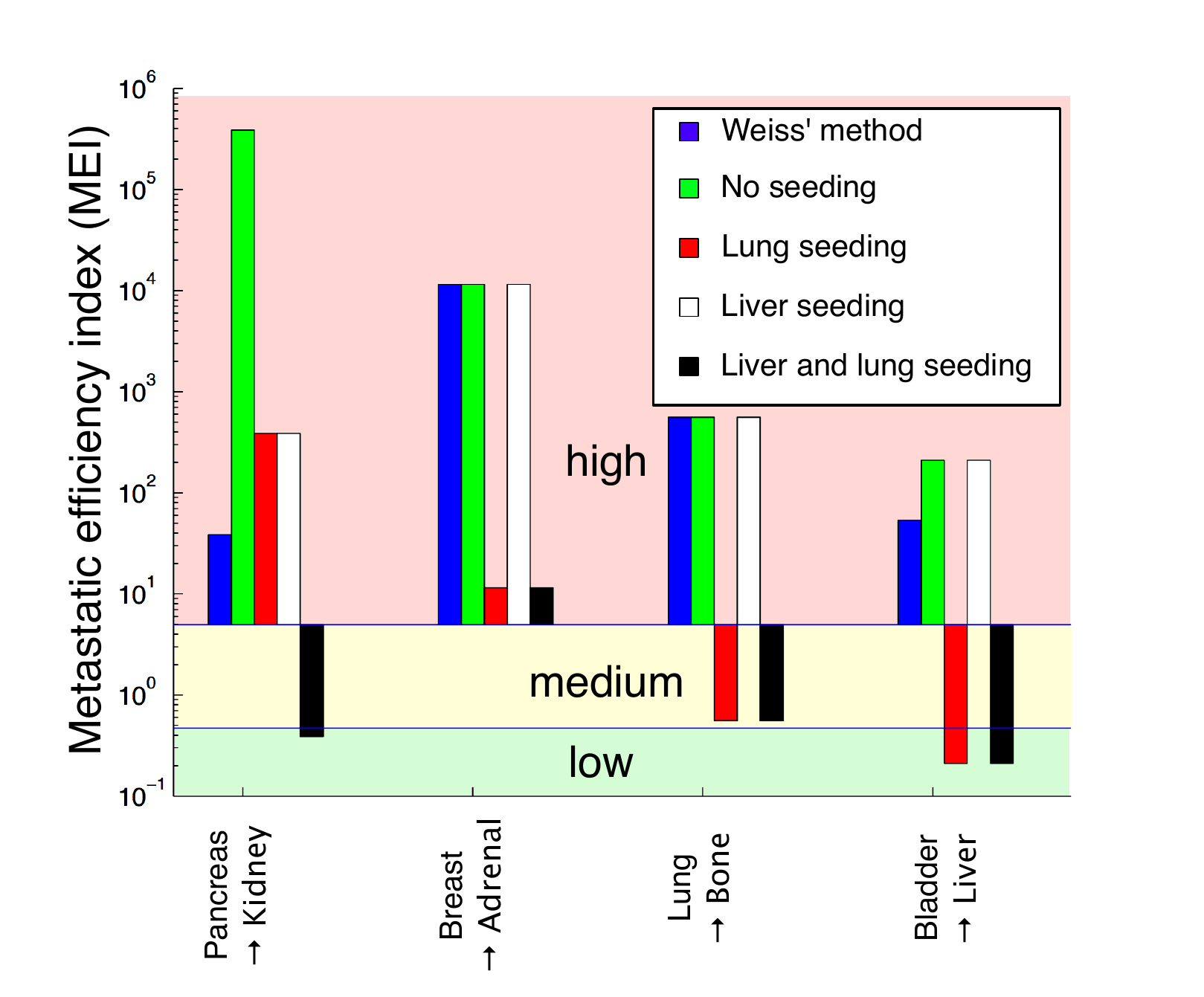}% Here is how to import EPS art
\caption{\label{fig:fig1}The impact of filter and flow characteristics on estimation of the metastatic efficienct index (MEI). We have compared Weiss' original method with our filter-flow framework under the assumption of no micrometastases, micrometastases in the lung, in the liver, and in both locations. The comparison is carried out for four organ pairs that cover the canonical pathways of spread (gut $\rightarrow$ body, body $\rightarrow$ body, lung $\rightarrow$ body and body $\rightarrow$ liver). We see that because Weiss' method only considers the dynamics on the arterial side it underestimates the MEI in two of the cases (pancreas $\rightarrow$ kidney, and bladder $\rightarrow$ liver). From the comparison it is also evident that assumptions about the presence or absence of micrometastases heavily influences the results, in the case of pancreas $\rightarrow$ kidney shifting the MEI six orders of magnitude from a low MEI to a high one (as defined by Weiss \cite{Weiss1992}).} 
\end{center}
\end{figure}

To illustrate the effect of micrometastatic disease on MEIs, we compare four scenarios for a set of representative organ pairs: no micrometastases, micrometastases present in the lung, in the liver and in both locations (Figure \ref{fig:fig1}). 
%
%We have compared Weiss' original method with MEIs calculated using the filter-flow framework in four different regimes: no micrometastases, micrometastases present in the lung, in the liver or in both locations. Figure \ref{fig:fig1} shows the result of this comparison for four organ pairs. 
We see that Weiss' metric differs from ours, but more importantly that the metastatic efficiency depends on the current disease state. For example, our estimate of the efficiency with which cells orginating from a primary pancreatic tumor can form kidney metastases varies over six orders of magnitude, depending on whether micrometastatic lesions are present, and their location. This effect highlights an opportunity to go a step further in disease characterization than presence or absence of CTCs at staging.

The preceding analysis assumes that the filtration rate for each organ is identical for all patients in the data set. This is likely a gross oversimplification, but no clinical trial has yet determined the intrapatient heterogeneity in this (currently absent) parameter set.
%In our filter-flow framework it is possible to view the incidence data as arising from a number of different patient groups, each following different routes of dissemination. 
Previously we used incidence data to calculate MEIs, but we may also reverse the process and calculate the prevalence of micrometastatic disease given incidence data and organ-pair MEIs.  Under this reasoning it is possible to show, for example, that the population incidence of metastases in the adrenal gland arising from primaries in the large intestine, which equals 7.5\%, can be explained not only by a single patient group composition, but by a whole collection (see online methods). Indeed, the same argument can be made for any primary-metastasis pair: the population-level data tell us nothing about a given patient.

%and it has previously been shown that most, but not all, patients with primary colon cancer follow the metachronous route of dissemination \cite{Weiss1992}.
%This type of analysis assumes that the MEI is constant across all patients, a statement which must remain controversial given our limited knowledge colonisation success, but one that will shed light on the patient-specific dynamics of metastatic spread. 

%The incidence can be explained by a subdivision according to 25 \% in the synchronous group, 20 \% in the liver seeding group, 5 \% in the lung seeding group, and 50\% of the patients exhibiting synchronous seeding, but also by a division of 5, 25, 20 and 50 \% into each patient group respectively. This highlights the fact that population-based measures of incidence can not be used to predict individual patient metastasis dynamics as different patients can exhibit fundamentally different patterns of spread. 

To enable these insights and their translation to the clinic, systematic testing of individual patient filter-flow parameters is required.  Measurement of CTCs at initial staging
%of surgery in each vascular compartment 
and subsequent correlation with outcomes would yield initial model parameters with which rational prospective trials could be designed.
%information that could lead to rational clinical trial design in patients with oligometastatic disease. 
This level of understanding of an individual patient's disease state constitutes a new type of personalized medicine, which seeks to assay not just the collection of mutations that a patient's cancer cells have accumulated, but instead their physical location. This would allow for more accurate staging and the rational inclusion of organ directed therapy in clinical trials, a concept which is gaining popularity with recently approved methods existing for bone and liver \citep{Harrison:2013vn,Seront:2012kx}.

By further elucidating the principles underlying hematogeonous metastasis, we hope to make inroads toward therapeutic strategy changes that would otherwise be impossible. Our results highlight the value of the physical perspective of the metastatic cascade, and the importance of addressing not only genetic factors, but also physiological and anatomical aspects of the process, which in this gene-centric era have been largely forgotten. 

\section*{Online methods}

\noindent \textbf{Calculation of Metastatic Efficiency Index (MEI)}\\
The autopsy dataset used in the analysis covers 3827 patients presenting with primary tumors in 30 different anatomical sites \cite{Disibio2008}. For each primary tumor the number of metastases are reported according to anatomical site (in total 9484 metastases). As we focus on the effect of blood flow patterns, we consider only the organs for which blood flow has been measured, and this reduces the number of anatomical sites to 14 (detailed in Table \ref{table:tab2}).

For each organ-organ pair we calculate the metastatic involvement $N_{ij}$ as the ratio between the number of occurrences of the primary tumor in organ $i$ and the number of metastases reported in the target organ $j$, for each pair $(i,j)$ presented in the autopsy data \cite{Disibio2008}. We have that $0 \leq N_{ij} < 1$ and this number corresponds to the fraction of cases where a primary tumor in organ $i$ gave rise to a metastasis in organ $j$. The metastatic efficiency index (MEI) from organ $i$ to $j$ is then defined by $M_{ij}=N_{ij}/\phi_{ij}$, where 
%$N_{ij}$ is the percent involvement of metastases in anatomical site $j$, given a primary at site $i$ (calculated from the autopsy data in \cite{Disibio2008}), and 
$\phi_{ij}$ is the relative flow of CTCs from organ $i$ to $j$. This quantity takes into account the blood flow that each target organ receives (Table \ref{table:tab1}) and the reduction in CTCs that occurs en route between the two organs. For the sake of simplicity we consider only the effects of capillary bed passage, and it has been shown in clinical studies that approximately 1 in 10 000 CTCs remain viable after such a passage \cite{Okumura:2009vn}. We thus assume that there occurs a reduction of CTC number by a factor $F_k$ when the cells pass through organ $k$. As a baseline, we use the pass rate $F_k=10^{-4}$ for all organs.

It is well known that metastases in the lung and liver have the ability to shed cells into the bloodstream and hence give rise to `second order' metastases \cite{Bross:1975p1937}. If one were to measure the CTC concentration downstream of an organ containing metastases, then it would be higher than in the case of a disease-free organ. For our purposes, this implies that the presence of metastatic disease can be represented in the model as a lower reduction of CTCs in the capillary bed of the affected organ. This simplification is only valid if we disregard the biological properties of the CTCs (since CTCs originating from metastases might have different genotypes and phenotypes compared to cells from the primary tumor), but is sufficient for our purposes. To simulate the presence of micrometastases in the lung and liver we therefore change the pass rates to $F_L = 10^{-1}$ and  $F_H = 10^{-1}$ respectively.

%As a baseline value we set $F_k=10^{-3}$

%which factors in a reduction in each capillary bed ($F_L$), and the relative blood flow to different organs as reported in  \cite{Williams:1989uq}. 

As an example of our methodology, we now present the calculations for the MEI for breast to adrenal gland. The cancer cells leaving a breast tumor enter the circulation on the venous side and are transported via the heart to the lung capillary bed, through which only a fraction $F_L$ pass as viable cells. These cells then flow into the arterial side of the circulation and are randomly distributed to the different organs of the body according to blood flow, of which the adrenal gland receives 0.3\% (Table \ref{table:tab1}). The relative flow of CTCs from breast to adrenal gland is therefore given by $\phi_{breast,adrenal}= F_L \times0.3 = 0.3 \times 10^{-3}$.% since the lung capillary bed filters out approximately 99.99\% of the cells and the relative blood flow to the adrenal gland is 0.3\%.

\vspace{1cm}\noindent \textbf{Patient group decomposition}\\ 
Although we calculated several distinct values of MEIs depending on the pattern of metastatic spread (presence of micrometastases), it is likely that most tumors originating in the same organ have roughly the same MEI, and that the presence of metastases in the liver and lung instead affect the incidence of secondary metastases. We now show how this can be used as a means to suggest possible patient group decompositions. 

The incidence, $N_{ij}$, relative flow of CTCs, $\phi_{ij}$ and the MEI, $M_{ij}$ are related according to $M_{ij}=N_{ij}/\phi_{ij}$, or equivalently $\phi_{ij} = N_{ij}/M_{ij}$. We now assume that $\phi_{ij}$ is not equal for all patients, and consider four patient groups: no micrometastases, micrometastases in the lung, micrometastases in the liver and micrometastases in both. If we now let $n_k$ denote the fraction of patients in each group, $k$, where $\sum n_k = 1$, then we can write
\begin{equation} \label{eq:groups}
\sum_{k=1}^4 n_k \phi_{ij}^k = M_{ij}/N_{ij},
\end{equation}
where $\phi_{ij}^k$ is the flow of CTCs in the different patient groups. This problem is underdetermined, and the solution (in terms of the fractions $n_i$) is given by any point on a surface defined by \eqref{eq:groups}, such that $n_i > 0$ for all patient groups and $\sum_i n_i=1$. This implies that aggregated incidence data can be explained by many different patient group compositions, each with its distinct pattern of metastatic progression. 

The population incidence of metastases in the adrenal gland arising from primaries in the large intestine equals 7.5 \%, and by fixing the MEI and using the above method, the incidence rate can be explained by a subdivision according to 25 \% in the no metastasis group, 20 \% in the liver metastases group, 5 \% in the lung metastases group, and 50\% of the patients harboring metastases in both liver and lung. However the incidence can also be explained by a subdivision of 5\%, 25\%, 20\% and 50 \% into each patient group respectively. This highlights the fact that population-based measures of incidence cannot be used to predict individual patient metastasis dynamics as different patients can exhibit fundamentally different patterns of spread. 

\begin{table}[htp]
\begin{center}
\caption{\label{table:tab1}The relative blood flow for the organs considered, taken from \cite{Williams:1989uq}. The model compartment refers to the anatomical location of the organs and their relation to the circulatory system is shown in fig. \ref{fig:filterflow}.}
    \vspace{0.5cm}
    \begin{tabular}{|l|l|l|}%{ | l | l | p{2cm} | p{3cm} |}
    \hline
    Tumor & Relative blood flow (\%) & Model compartment \\%& No.\ of primaries in organ & No.\ of mets in organ \\ 
    \hline
       Liver & 6.5 (arterial) + 19 (portal) & H\\%& 36 & 760 \\
       Lung & 100 & L\\%& 136 & 695 \\
       Adrenal & 0.3 & B\\%& 6 & 383 \\
       Bladder & 0.06 & B\\%& 183 & 42 \\
       Bone & 5.0 & B\\%& 35 & 426 \\
       Breast & 1.0 & B\\%& 432 & 61 \\
       Large intestine & 4.0 & G\\%& 560 & 61 \\
       Kidney & 19.0 & B\\%& 62 & 162 \\
       Pancreas & 1.0 & B\\%& 109 & 168 \\
       Skin & 5.0 & B\\%& 161 & 197 \\
       Small intestine & 10.0 & G\\%& 19 & 80 \\
       Stomach  & 1.0 & G\\%& 477 & 41 \\
       Testes & 0.05 & B\\%& 25 & 5 \\
       Thyroid & 1.5 & B\\%& 43 & 89 \\
       \hline
    \end{tabular}
    \end{center}
\end{table}

\begin{table}[htp]
\begin{center}
\small
\caption{\label{table:tab2}Distribution of metastases according to primary site. Data taken from \cite{Disibio2008}.}
    \vspace{0.5cm}

\begin{tabular}{|l|c|cccccccccccccc|c|c|}\hline
Primary site& No. of cases
%\backslashbox{Room}{Date}
&\begin{sideways}Liver\end{sideways}
&\begin{sideways}Lung\end{sideways}
&\begin{sideways}Adrenal\end{sideways}
&\begin{sideways}Bladder\end{sideways}
&\begin{sideways}Bone\end{sideways}
&\begin{sideways}Breast\end{sideways}
&\begin{sideways}Large intestine\end{sideways}
&\begin{sideways}Kidney\end{sideways}
&\begin{sideways}Pancreas\end{sideways}
&\begin{sideways}Skin\end{sideways}
&\begin{sideways}Small intestine\end{sideways}
&\begin{sideways}Stomach\end{sideways}
&\begin{sideways}Testes\end{sideways}
&\begin{sideways}Thryoid\end{sideways}

\\ \hline
Liver & 36 & 0 &16 &7 &0 &3 &0 &1& 0& 4& 1& 1& 0& 1& 1  \\
Lung & 136 & 58& 48& 59& 1 &38& 0& 9& 30& 23& 8 &6& 5& 0& 15\\
Adrenal &6 &3 &2 &1 &0 &3 &0 &1 &1 &1 &3 &0 &1 &0 &1\\
Bladder &183 &25& 30& 11& 0& 20& 0& 1& 9& 2& 1& 4& 0& 0& 1\\
Bone &35 & 6 &18 &4 &1 &15& 0 &1 &2 &3& 6& 0& 0& 0& 0\\
Breast & 432 & 218 &247 &149 &17 &213 &54 &11& 40& 49 &124 &12 &17 &0 &35\\
Large intestine &560 & 155 & 101 & 42 & 10 & 29 & 0 & 8 & 20 & 13 & 9 & 6 & 1 & 0 &8\\
Kidney & 62 &21 &30 &18 &2& 20& 1 &5 &8& 9& 3& 4& 2& 0& 1\\
Pancreas & 109 &63 &29 &12 &2 &7& 1& 4 &8 &0& 3& 6& 4& 1& 3\\
Skin &161 &28& 47 &25 &2 &25& 3& 3& 18& 14& 22 &11 &5 &2 &14\\
Small intestine & 19 &8& 3& 0 &0 &0 &0 &2& 0& 0& 0& 1& 1& 0& 0\\
Stomach & 477 &146& 84& 45& 7 &39& 2& 13& 14& 44 &11 &25& 3& 0& 5\\
Testes & 25 &19 &18 &6 &0 &8 &0 &2 &11 &3& 2& 3& 2& 1& 2\\
Thyroid & 43 &10& 22& 4& 0 &6 &0 &0 &1& 3& 4& 1& 0& 0& 0\\
\hline
%\multicolumn{2}{|c|}{Total } & 695 & 383 & 42 & 426 & 61 & 61 & 162 & 168 & 197 & 80 & 41 & 5 & 89  \\
%\hline
\end{tabular}
\end{center}
\end{table}
\bibliography{mets.bib}

\begin{thebibliography}{20}
\providecommand{\natexlab}[1]{#1}
\providecommand{\url}[1]{\texttt{#1}}
\expandafter\ifx\csname urlstyle\endcsname\relax
  \providecommand{\doi}[1]{doi: #1}\else
  \providecommand{\doi}{doi: \begingroup \urlstyle{rm}\Url}\fi

\bibitem[Ashworth(1869)]{Ashworth}
T.R. Ashworth.
\newblock A case of cancer in which cells similar to those in the tumours were
  seen in the blood after death.
\newblock \emph{Australian Medical Journal}, 14\penalty0 (3-4):\penalty0
  146--147, 1869.

\bibitem[Plaks et~al.(2013)Plaks, Koopman, and Werb]{Plaks}
V~Plaks, C.~D Koopman, and Z~Werb.
\newblock Circulating tumor cells.
\newblock \emph{Science}, 341\penalty0 (6151):\penalty0 1186--1188, Sep 2013.
\newblock \doi{10.1126/science.1235226}.

\bibitem[Paget(1989)]{Paget:1989ys}
S.~Paget.
\newblock The distribution of secondary growths in cancer of the breast. 1889.
\newblock \emph{Cancer Metastasis Rev.}, 8\penalty0 (2):\penalty0 98--101,
  1989.

\bibitem[Rapp(2001)]{weissmemoriam}
D.G. Rapp.
\newblock In memoriam {Leonard} {L.} {Weiss}, {Sc}.{D}., {M}.{D}., {Ph}.{D}.
\newblock \emph{Cancer Res.}, 61:\penalty0 5663, 2001.

\bibitem[Weiss(1992)]{Weiss1992}
L.~Weiss.
\newblock Comments on hematogenous metastatic patterns in humans as revealed by
  autopsy.
\newblock \emph{Clin. Exp. Metastasis}, 10\penalty0 (3):\penalty0 191--199,
  1992.

\bibitem[Minn et~al.(2005)Minn, Gupta, Siegel, Bos, Shu, Giri, Viale, Olshen,
  Gerald, and Massagu{\'e}]{Minn:2005ia}
A.J. Minn, G.P. Gupta, P.M. Siegel, P.D. Bos, W.~Shu, D.D. Giri, A.~Viale, A.B.
  Olshen, W.L. Gerald, and J.~Massagu{\'e}.
\newblock {Genes that mediate breast cancer metastasis to lung}.
\newblock 436\penalty0 (7050):\penalty0 518--524, 2005.

\bibitem[Bos et~al.(2009)Bos, Zhang, Nadal, Shu, Gomis, Nguyen, Minn, van~de
  Vijver, Gerald, Foekens, and Massagu{\'e}]{Bos:2009jl}
P.D. Bos, X.H.F. Zhang, C.~Nadal, W.~Shu, R.R. Gomis, D.X. Nguyen, A.J. Minn,
  M.J. van~de Vijver, W.L. Gerald, J.A. Foekens, and J.~Massagu{\'e}.
\newblock {Genes that mediate breast cancer metastasis to the brain.}
\newblock \emph{Nature}, 459\penalty0 (7249):\penalty0 1005--1009, 2009.

\bibitem[Marusyk et~al.(2012)Marusyk, Almendro, and Polyak]{Marusyk:2012uq}
A.~Marusyk, V.~Almendro, and K.~Polyak.
\newblock Intra-tumour heterogeneity: a looking glass for cancer?
\newblock \emph{Nat. Rev. Cancer}, 12\penalty0 (5):\penalty0 323--34, 2012.
\newblock \doi{10.1038/nrc3261}.

\bibitem[Gerlinger et~al.(2012)Gerlinger, Rowan, Horswell, Larkin, Endesfelder,
  Gronroos, Martinez, Matthews, Stewart, Tarpey, Varela, Phillimore, Begum,
  McDonald, Butler, Jones, Raine, Latimer, Santos, Nohadani, Eklund,
  Spencer-Dene, Clark, Pickering, Stamp, Gore, Szallasi, Downward, Futreal, and
  Swanton]{Gerlinger:2012uq}
M.~Gerlinger, A.J. Rowan, S.~Horswell, J.~Larkin, David Endesfelder,
  E.~Gronroos, P.~Martinez, N.~Matthews, A.~Stewart, P.~Tarpey, I.~Varela,
  B.~Phillimore, S.~Begum, N.Q. McDonald, A.~Butler, D.~Jones, K.~Raine,
  C.~Latimer, C.R. Santos, M.~Nohadani, A.C. Eklund, B.~Spencer-Dene, G.~Clark,
  L.~Pickering, G.~Stamp, M.~Gore, Z.~Szallasi, J.~Downward, P.A. Futreal, and
  C.~Swanton.
\newblock Intratumor heterogeneity and branched evolution revealed by
  multiregion sequencing.
\newblock \emph{N. Engl. J. Med.}, 366\penalty0 (10):\penalty0 883--92, 2012.
\newblock \doi{10.1056/NEJMoa1113205}.

\bibitem[Scott et~al.(2012)Scott, Kuhn, and Anderson]{Scott:2012p1778}
J.G. Scott, P.~Kuhn, and A.R.A Anderson.
\newblock Unifying metastasis --- integrating intravasation, circulation and
  end-organ colonization.
\newblock \emph{Nat. Rev. Cancer}, 12:\penalty0 1--2, 2012.
\newblock \doi{10.1038/nrc3287}.
\newblock URL \url{http://dx.doi.org/10.1038/nrc3287}.

\bibitem[Scott et~al.(2013)Scott, Basanta, Anderson, and
  Gerlee]{Scott:2013p1914}
J.G. Scott, D.~Basanta, A.R.A. Anderson, and P.~Gerlee.
\newblock A mathematical model of tumour self-seeding reveals secondary
  metastatic deposits as drivers of primary tumour growth.
\newblock \emph{J. R. Soc. Interface}, 10:\penalty0 1--10, 2013.

\bibitem[Milano et~al.(2009)Milano, Zhang, Metcalfe, Muhs, and
  Okunieff]{Milano:2009zr}
M.~T. Milano, H.~Zhang, S.K. Metcalfe, A.G. Muhs, and P.~Okunieff.
\newblock Oligometastatic breast cancer treated with curative-intent
  stereotactic body radiation therapy.
\newblock \emph{Breast Cancer Res. Treat.}, 115\penalty0 (3):\penalty0 601--8,
  2009.
\newblock \doi{10.1007/s10549-008-0157-4}.

\bibitem[Milano et~al.(2012)Milano, Katz, Zhang, and Okunieff]{Milano:2012ys}
M.T. Milano, A.W. Katz, H.~Zhang, and P.~Okunieff.
\newblock Oligometastases treated with stereotactic body radiotherapy:
  long-term follow-up of prospective study.
\newblock \emph{Int. J. Radiat. Oncol. Biol. Phys.}, 83\penalty0 (3):\penalty0
  878--86, 2012.
\newblock \doi{10.1016/j.ijrobp.2011.08.036}.

\bibitem[Weichselbaum and Hellman(2011)]{Weichselbaum:2011uq}
R.R. Weichselbaum and S.~Hellman.
\newblock Oligometastases revisited.
\newblock \emph{Nat. Rev. Clin. Oncol.}, 8\penalty0 (6):\penalty0 378--82,
  2011.
\newblock \doi{10.1038/nrclinonc.2011.44}.

\bibitem[Williams and Leggett(1989)]{Williams:1989uq}
L.R. Williams and R.W. Leggett.
\newblock Reference values for resting blood flow to organs of man.
\newblock \emph{Clin. Phys. Physiol. Meas.}, 10\penalty0 (3):\penalty0
  187--217, 1989.

\bibitem[Disibio and French(2008)]{Disibio2008}
G.~Disibio and S.W. French.
\newblock Metastatic patterns of cancers: results from a large autopsy study.
\newblock \emph{Arch. Pathol. Lab. Med.}, 132\penalty0 (6):\penalty0 931--939,
  2008.
\newblock \doi{10.1043/1543-2165(2008)132[931:MPOCRF]2.0.CO;2}.
\newblock URL
  \url{http://dx.doi.org/10.1043/1543-2165(2008)132[931:MPOCRF]2.0.CO;2}.

\bibitem[Okumura et~al.(2009)Okumura, Tanaka, Yoneda, Hashimoto, Takuwa, Kondo,
  and Hasegawa]{Okumura:2009vn}
Y.~Okumura, F.~Tanaka, K.~Yoneda, M.~Hashimoto, T.~Takuwa, N.~Kondo, and
  S.~Hasegawa.
\newblock Circulating tumor cells in pulmonary venous blood of primary lung
  cancer patients.
\newblock \emph{Ann. Thorac. Surg.}, 87\penalty0 (6):\penalty0 1669--75, 2009.
\newblock \doi{10.1016/j.athoracsur.2009.03.073}.

\bibitem[Harrison et~al.(2013)Harrison, Wong, Armstrong, and
  George]{Harrison:2013vn}
M.R. Harrison, T.Z. Wong, A.J. Armstrong, and D.J. George.
\newblock Radium-223 chloride: a potential new treatment for
  castration-resistant prostate cancer patients with metastatic bone disease.
\newblock \emph{Cancer Manag. Res.}, 5:\penalty0 1--14, 2013.
\newblock \doi{10.2147/CMAR.S25537}.

\bibitem[Seront and Van~den Eynde(2012)]{Seront:2012kx}
E.~Seront and M.~Van~den Eynde.
\newblock Liver-directed therapies: does it make sense in the current
  therapeutic strategy for patients with confined liver colorectal metastases?
\newblock \emph{Clin. Colorectal Cancer}, 11\penalty0 (3):\penalty0 177--84,
  2012.
\newblock \doi{10.1016/j.clcc.2011.12.004}.

\bibitem[Bross et~al.(1975)Bross, Viadana, and Pickren]{Bross:1975p1937}
I.~Bross, W.~Viadana, and J.~Pickren.
\newblock Do generalized metastases occur directly from the primary?
\newblock \emph{J. Chronic Dis.}, 28\penalty0 (3):\penalty0 149--159, 1975.

\end{thebibliography}
\bibliographystyle{unsrtnat}
%\bibliographystyle{unsrt}
%\bibliographystyle{plain}
%\bibliography{citations_version7}

\end{document}